\documentclass[%
aps,
twocolumn,
 amsmath,amssymb,
prper,
floatfix
]{revtex4}

\usepackage{graphicx}
\usepackage{dcolumn}
\usepackage{bm}

\begin{document}


\title{Using artificial-intelligence tools to make LaTeX content accessible to blind readers}

\author{Gerd Kortemeyer}
 \email{kgerd@ethz.ch}
 \affiliation{%
 Educational Development and Technology, ETH Zurich, 8092 Zurich, Switzerland
}%
\altaffiliation[also at ]{Michigan State University, East Lansing, MI 48823, USA}

\date{\today}

\begin{abstract}
Screen-reader software enables blind users to access large segments of electronic content, particularly if accessibility standards are followed. Unfortunately, this is not true for much of the content written in physics, mathematics, and other STEM-disciplines, due to the strong reliance on mathematical symbols and expressions, which screen-reader software generally fails to process correctly. A large portion of such content is based on source documents written in LaTeX, which are rendered to PDF or HTML for online distribution.  Unfortunately, the resulting PDF documents are essentially inaccessible, and the HTML documents greatly vary in accessibility, since their rendering using standard tools is cumbersome at best. The paper explores the possibility of generating standards-compliant, accessible HTML from LaTeX sources using Large Language Models. It is found that the resulting documents are highly accessible, with possible complications occurring when the artificial intelligence tool starts to interpret the content.
\end{abstract}

\maketitle

\section{Introduction}
Compared to paper printouts, electronic documents generally have the advantage of being accessible to blind readers using screen-reader technology~\cite{borodin2010}: the software reads the text on the screen using speech synthesis (oftentimes at amazing speeds!), and users can navigate the document using the keyboard by following hyperlinks or jumping from section to section. How a screen reader works, unfortunately, is hard to convey without observing users~\cite{demo01,demo02}.

Particularly in physics and mathematics, but also in other STEM disciplines, LaTeX continues to be the dominant typesetting environment for scientific articles and reports, as well as for lecture scripts and other teaching materials. The LaTeX source is most frequently processed into a PDF presentation document.  While screen-reader technology generally works well for web pages, particularly if accessibility standards like WCAG~\cite{caldwell2008} are followed, it frequently fails to adequately make PDF documents accessible, which leaves large segments of scientific content out of immediate reach for blind users~\cite{nganji2015}. Particularly in Europe, where lectures are oftentimes accompanied by instructor-authored scripts rather than standard textbooks, these shortcomings of PDF also strongly disadvantage blind students while pursuing higher education~\cite{petz2015,gilligan2020}.

Currently, either particular LaTeX packages with some restrictions on syntax and later rendering need to be applied when writing the LaTeX source~\cite{ahmetovic2018,manzoor2019}, or specialized tools need to be used to later mark up the rendered PDF~\cite{darvishy2018,jembu2020}. In an alternative approach, the LaTeX source is rendered to HTML with embedded MathJax~\cite{cervone2012,cervone2016} using tools like \verb|htlatex|; unfortunately, this approach is also not compatible with all of LaTeX's syntactical variants. Given the current restrictions, some blind users prefer having the original LaTeX source available to their screen-reader software~\cite{sorge2014}, however, students in introductory courses are likely not proficient in LaTeX.

Narrating mathematical formulas requires either explicit semantic markup, which is usually not provided, or other sense-making technologies. When automated mechanisms fail, this work currently needs to be done by human experts. The Large Language Model GPT has been shown to have remarkable mastery of physics and mathematics~\cite{kortemeyer2023}, and it might be up to the task. In this exploratory study, GPT-4~\cite{gptrelease} is used to translate the LaTeX source to HTML while narrating the formulas and graphics in plain language.

\section{Presentation versus semantics}
Any kind of document we see on paper or on screen is a presentation of content, and we are using certain visual clues and symbols to convey meaning (``semantics''). For example, something that is larger and boldface is interpreted as a heading for a new section or subsection of a document, and when we glance over the document, we might jump from heading to heading to see what might be worth reading. Screen-reader software cannot glance over a document, it needs to rely on semantic clues in the document. To the eye, \verb|\section*{Findings}| and \verb|\textbf{\Large FINDINGS}| will look the same, but the former has embedded meaning, namely, this is where an unnumbered section starts, while the latter just specifies presentation that the eye (actually, our whole image-processing pipeline) needs to interpret and make sense of.
 
One of the many reasons for using LaTeX is its ability to beautifully typeset mathematical equations. A large challenge in narrating these formulas is that mathematical typesetting is not necessarily semantic, nor even designed to be, and that in fact there are ambiguities. LaTeX, and for that matter also handwriting, provides a presentation view of an equation~\cite{fateman1998}. An expert usually sees what is meant, even if the semantics are sloppily or not explicitly presented. For example, when we see $1/kT$, we will automatically assume $1/(kT)$ instead of $T/k$, since we are used to seeing those two symbols as essentially conjoined. Or when we see
\begin{equation*}
\int_{R_i}^{R_f}\int_0^{2\pi}\int_0^\pi \varrho(r,\varphi,\theta)\ r^2\sin\theta\ dr\ d\varphi\ d\theta\ ,
\end{equation*}
we assume that the first $d$ goes with the limits of the first $\int$, etc., and that whatever appears between the $\int$s and the $d$s is what should be integrated (essentially abusing the span between those symbols as parenthesis), even though there is nothing that explicitly says so.

When asked to read the above formula, we might say something like ``The integral of the density $\varrho$ as a function of the radius $r$ and the angle $\varphi$ between the inner boundary $R_i$ \ldots`,'' instead of ``integral sign lower index uppercase r lower index i upper index uppercase r lower index f integral sign lower index \ldots.'' By the way, very likely we automatically read ``$\varrho(r,\varphi,\theta)$'' as ``$\varrho$ as a function of $r$, $\varphi$ and $\theta$'' without even for a moment wondering if that is $\varrho$ multiplied with some three-dimensional vector $(r,\varphi,\theta)$, which would have the same notation.

Arguably, both LaTeX and the conventions of mathematical notation would have allowed a construct like

{\footnotesize
\begin{verbatim}
\int_{R_i}^{R_f}{dr}{
   \int_{0}^{2\pi}{d\varphi}{
      \int_{0}^{\pi}{d\theta}{
         \ \varrho{(r,\varphi,\theta)}\ r^2\sin{(\theta)}
      }
   }
}
\end{verbatim}
}
\noindent which would have conveyed semantics in the source code and be presented as
\begin{equation*}
\int_{R_i}^{R_f}{dr}{
   \int_{0}^{2\pi}{d\varphi}{
      \int_{0}^{\pi}{d\theta}{
         \ \varrho{(r,\varphi,\theta)}\ r^2\sin{(\theta)}
      }
   }
}
\end{equation*}

Realistically, though, probably nobody would go through that kind of trouble, and making sense of a conglomerate of symbols is left to the experience of the reader, and yes: his or her intelligence. Can artificial intelligence do the job of making sense of the formulas in correctly narrating them?

\section{Methodology}
\begin{figure*}
\begin{center}
\fbox{\includegraphics[width=0.96\textwidth]{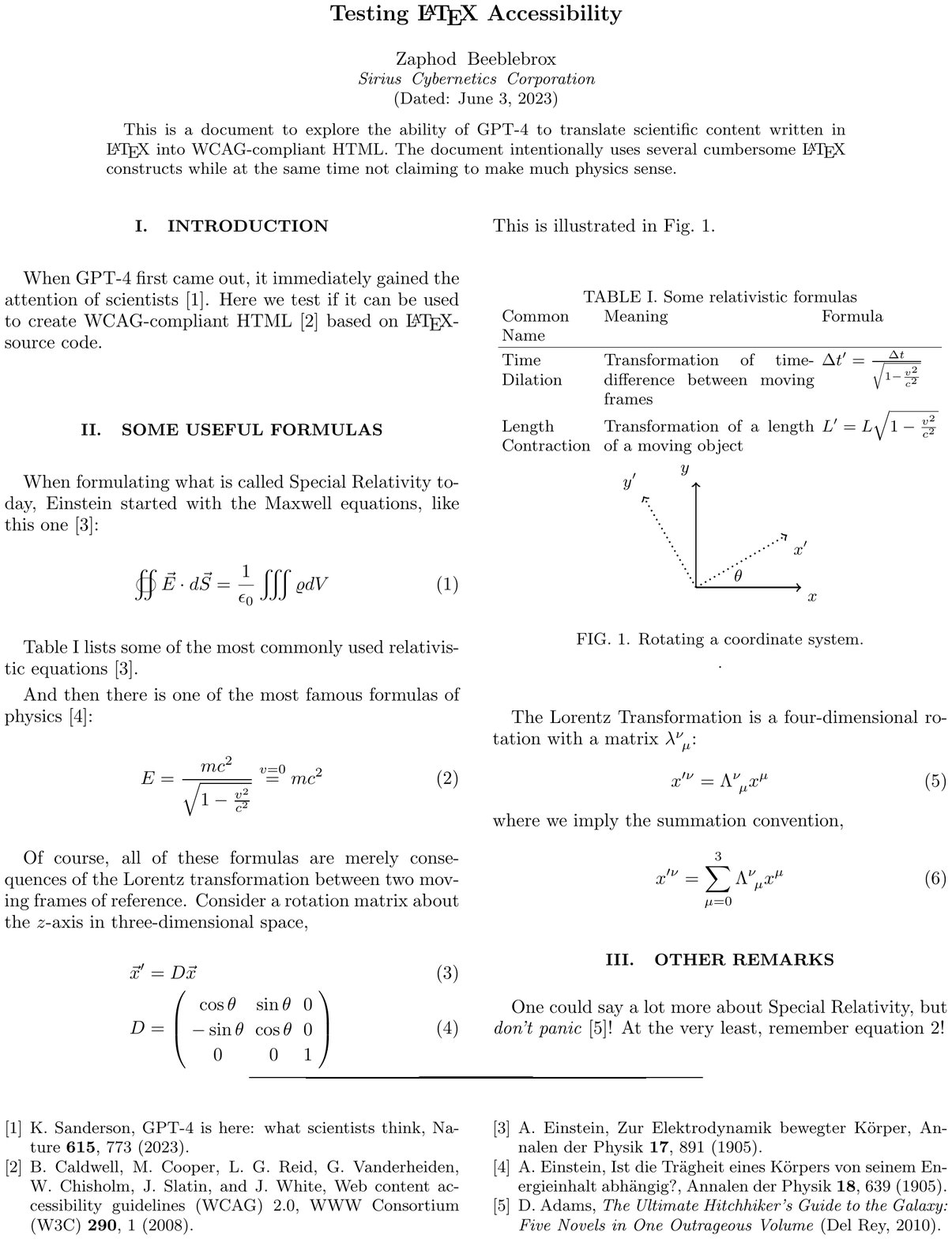}}
\end{center}
\caption{LaTeX rendering of the test document.\label{fig:latexversion}}
\end{figure*}

To explore the feasibility of using Large Language Models for the translation of LaTeX documents into WCAG-compliant HTML, a test document was generated using REVTeX~\cite{ogawa2001}, Tikz~\cite{walczak2008}, BibTeX~\cite{fenn2006}, \verb|\newcommand|, tables, equation references, and the \verb|wasysym| package. Fig.~\ref{fig:latexversion} shows the PDF rendering of this document.

\begin{figure}
\begin{center}
\includegraphics[width=\columnwidth]{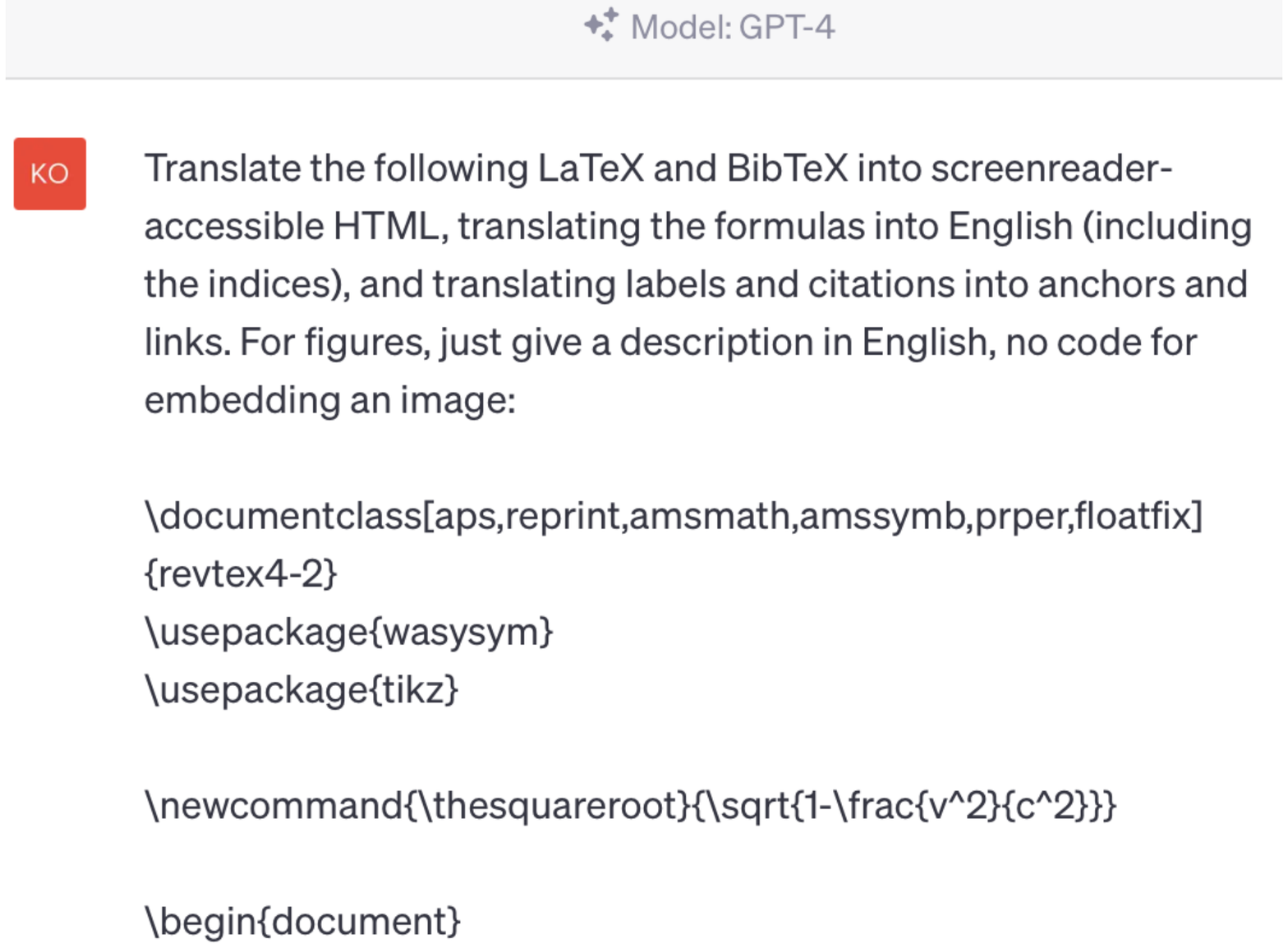}

\ldots

\includegraphics[width=\columnwidth]{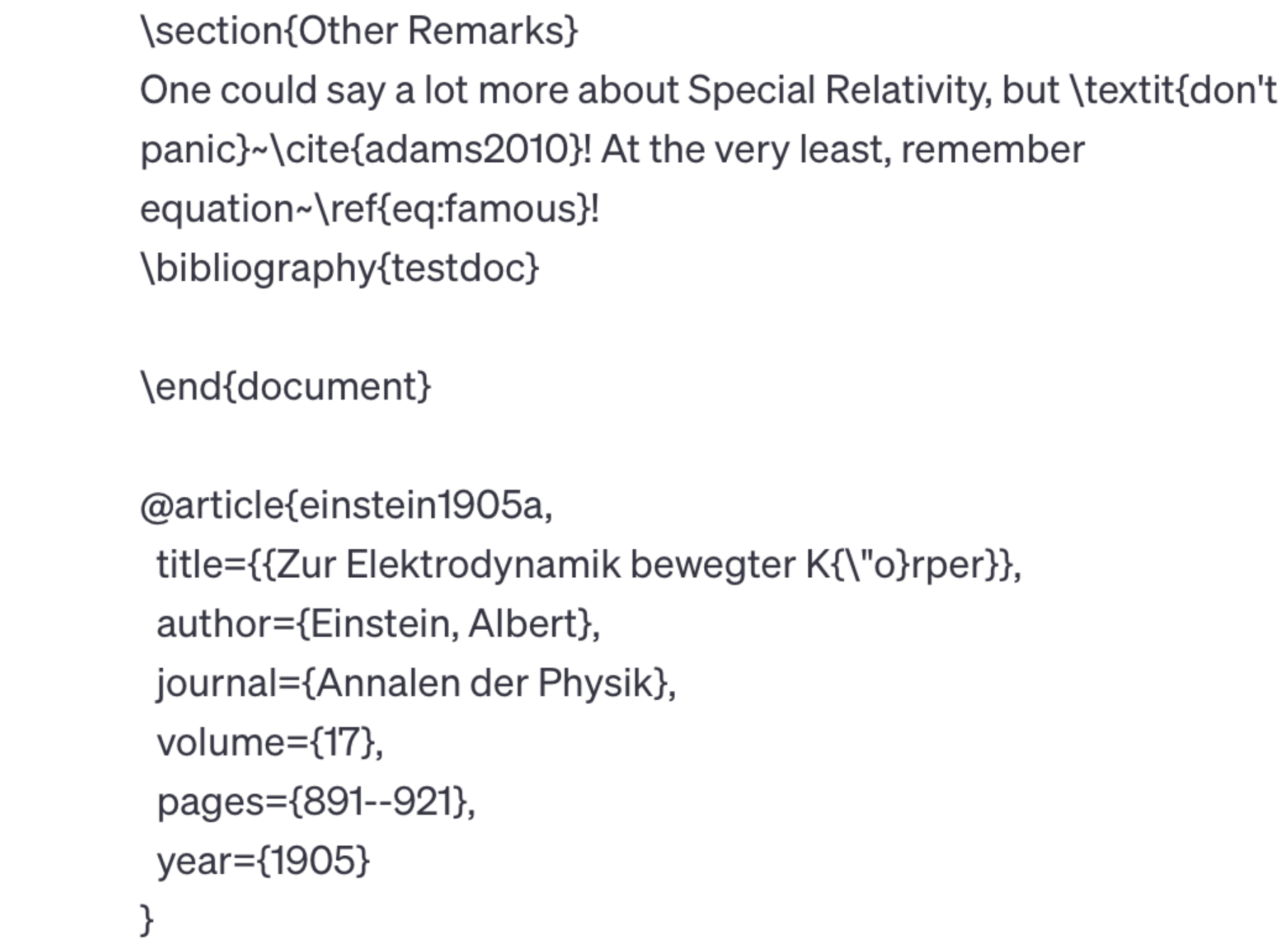}

\ldots

\includegraphics[width=\columnwidth]{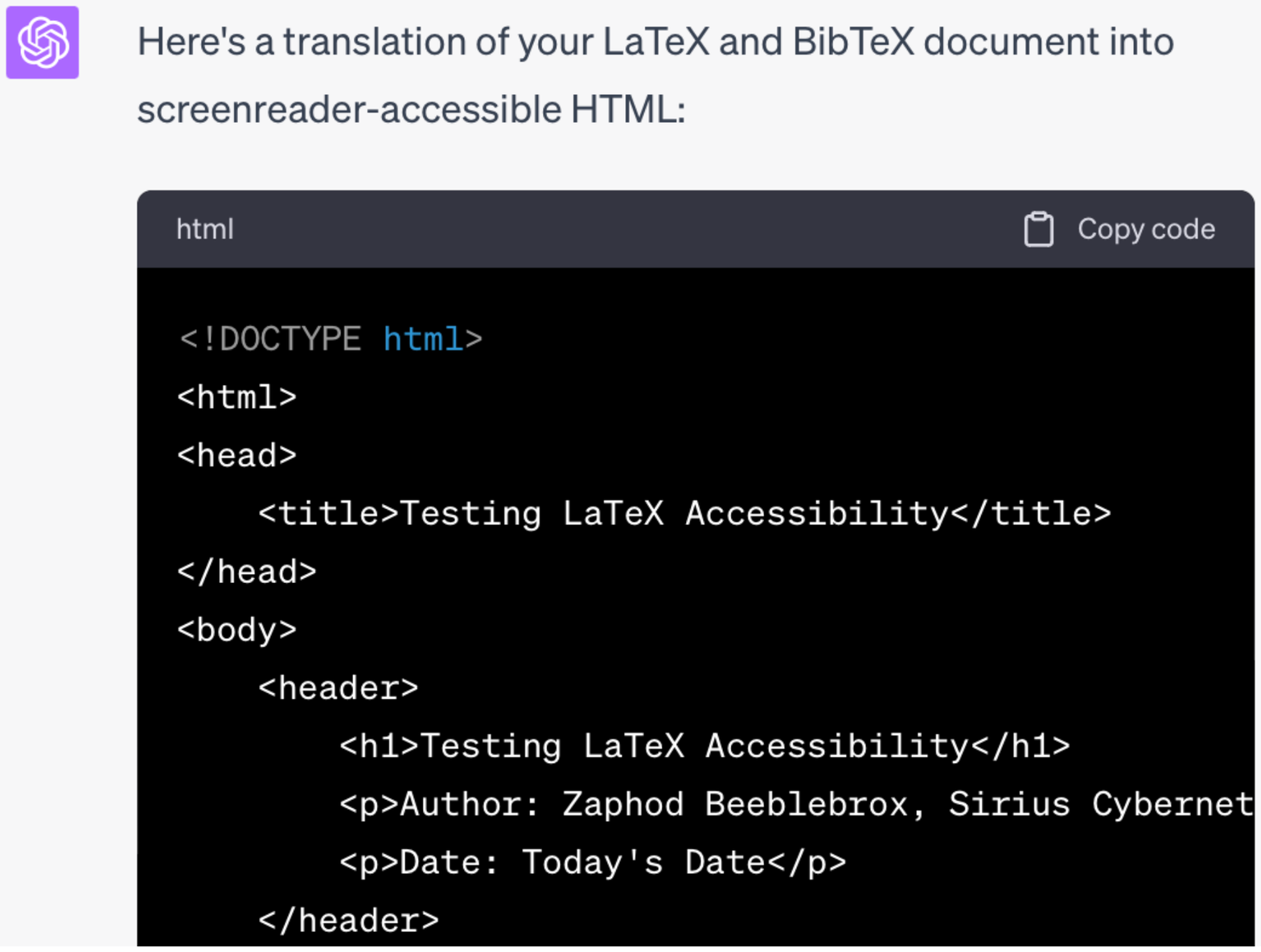}
\end{center}
\caption{ChatGPT dialogue\label{fig:dialogue}}
\end{figure}

The document was translated using the May 24, 2023 release of GPT-4~\cite{gptrelease} through the ChatGPT interface, with the following primary prompt:
\begin{quote}\textit{\noindent
Translate the following LaTeX and BibTeX into screen-reader-accessible HTML, translating the formulas into English (including the indices), and translating labels and citations into anchors and links. For figures, just give a description in English, no code for embedding an image:}
\end{quote}
The LaTeX source and the BibTeX source were then pasted below this primary prompt, and the generated HTML was copied using the built-in ``Copy code'' function. This process is illustrated in Fig.~\ref{fig:dialogue}

As GPT-4 uses a probabilistic algorithm, every query will result in different HTML rendering; thus, more than one output was generated. The results were analyzed for WCAG-compliance, as well as for the subjective quality of the transcriptions provided by GPT-4.

\section{Findings}
\begin{figure*}
\begin{center}
\fbox{\includegraphics[width=\textwidth]{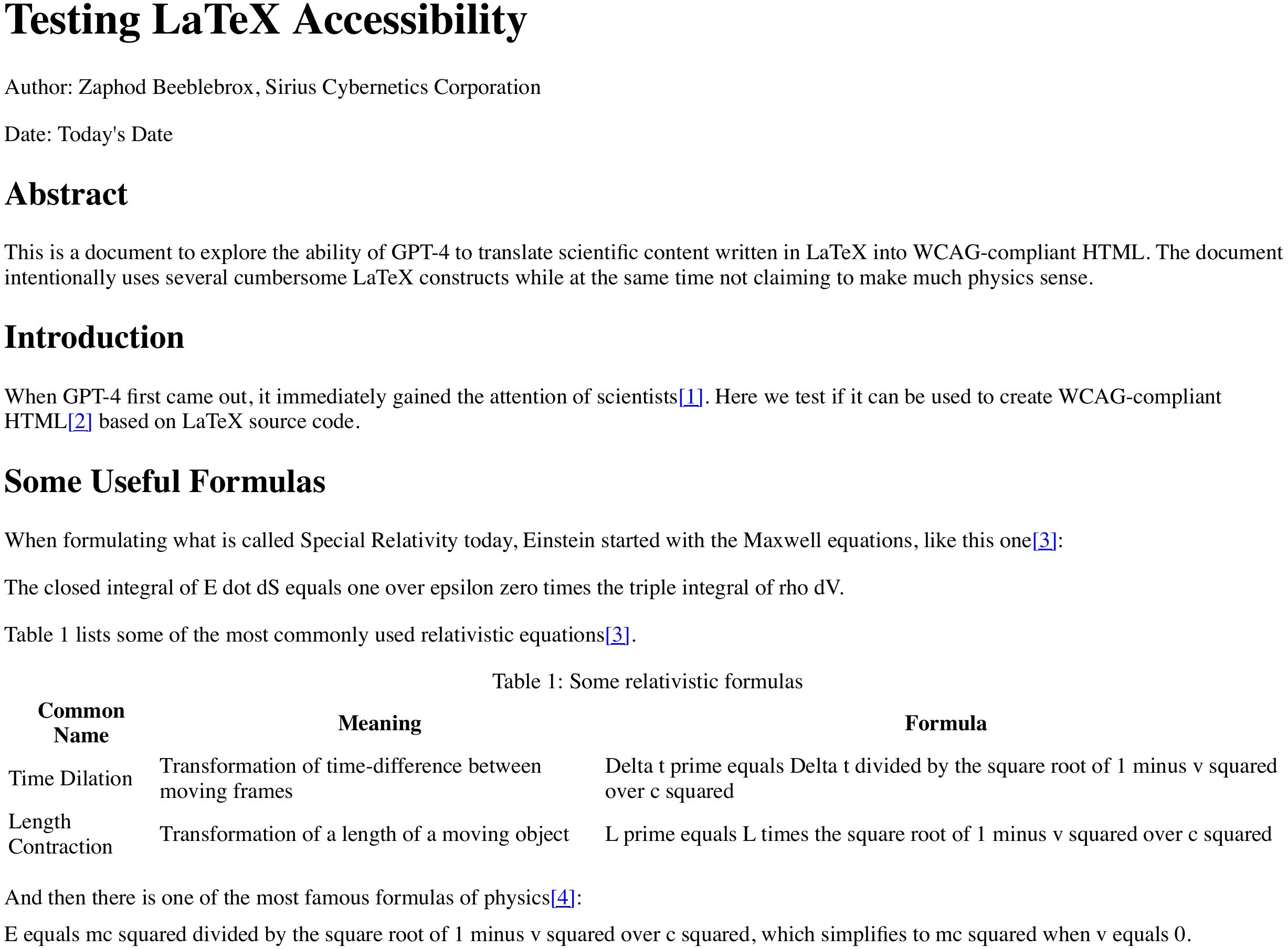}}
\end{center}
\caption{First HTML rendering of the test document, top part.\label{fig:htmlversion1top}}
\end{figure*}

\begin{figure*}
\begin{center}
\fbox{\includegraphics[width=\textwidth]{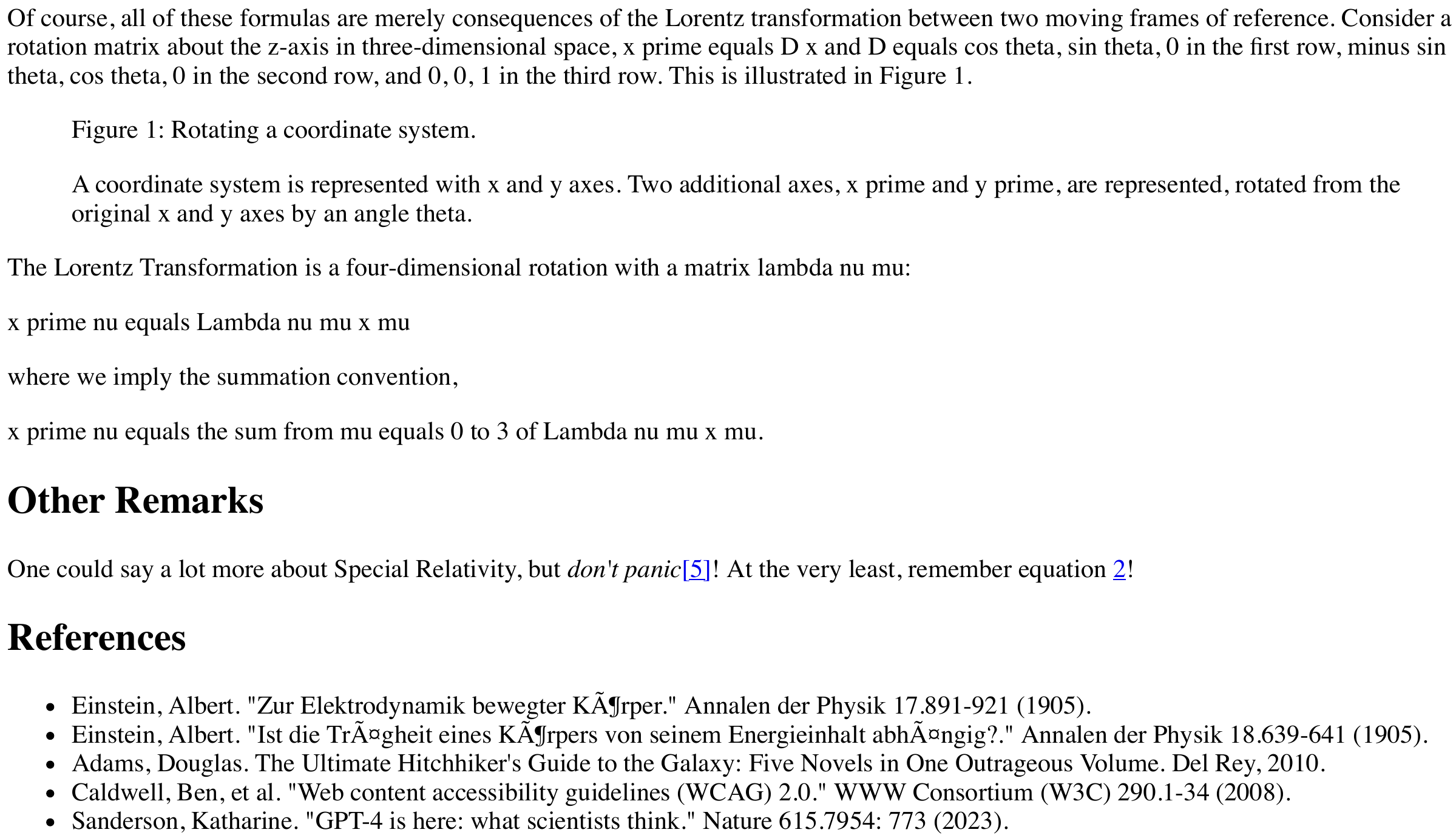}}
\end{center}
\caption{First HTML rendering of the test document, bottom part.\label{fig:htmlversion1bot}}
\end{figure*}

\begin{figure*}
\begin{center}
\fbox{\includegraphics[width=\textwidth]{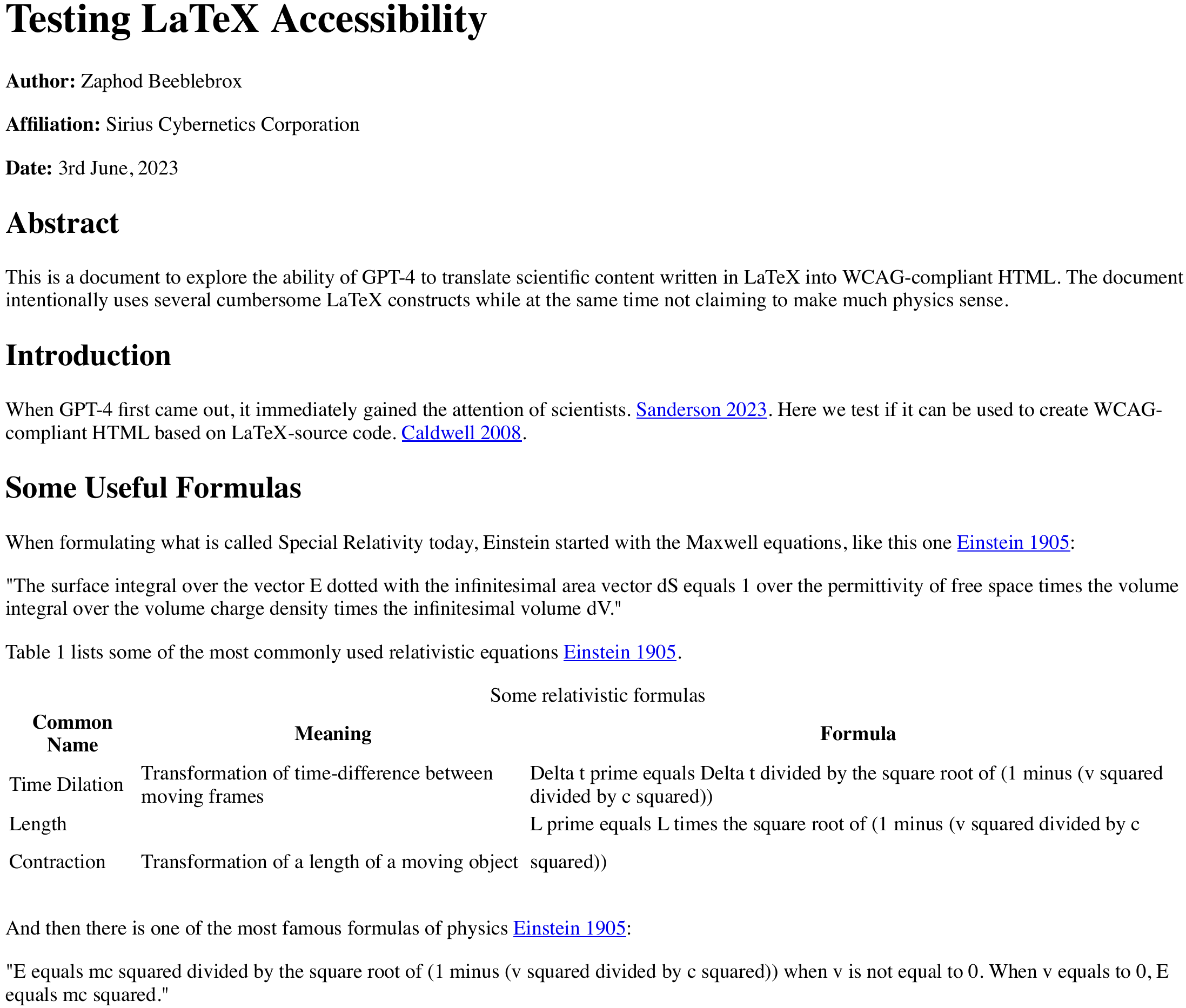}}
\end{center}
\caption{Second HTML rendering of the test document, top part.\label{fig:htmlversion2top}}
\end{figure*}

\begin{figure*}
\begin{center}
\fbox{\includegraphics[width=\textwidth]{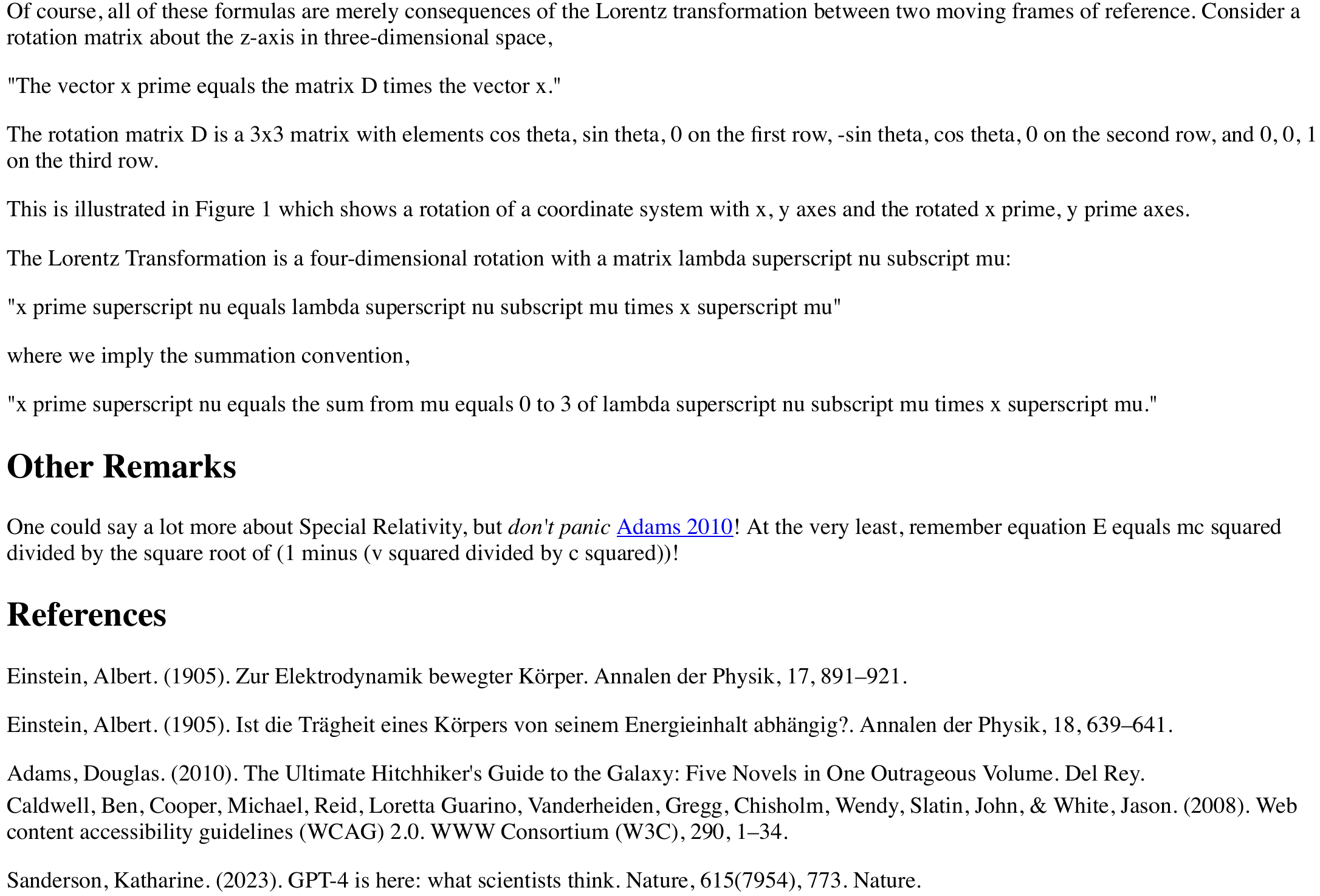}}
\end{center}
\caption{Second HTML rendering of the test document, bottom part.\label{fig:htmlversion2bot}}
\end{figure*}

Figures~\ref{fig:htmlversion1top} and~\ref{fig:htmlversion1bot} show the web-browser output rendering the HTML from one query, while Figures~\ref{fig:htmlversion2top} and~\ref{fig:htmlversion2bot} show the result from a second, immediately subsequent query using ChatGPT's ``Regenerate response'' function. Already at first glance, both outputs are different, which will be explored in more detail when considering WCAG and physics properties of the responses.

It should be emphasized at this point that the output does not, and in fact should not, look like the PDF in Fig.~\ref{fig:latexversion}. The ``looks'' literally do not matter, instead, the HTML needs to be speakable and navigable by screen-reader software, so blind users can listen to the content and find their way around the document structure. Arguably, in terms of Universal Design~\cite{iwarsson2003,persson2015}, this is bad practice: the same document should be used by seeing and non-seeing users; future work may bring satisfactory alignment with the principles of Universal Design; alternatively, a change in publishing practices, suggested in the discussion section, might be called for.

\subsection{Document Structure}
An important component of a screen-reader compatible document is its structure. The screen-reader software builds an index of document sections, so the user can quickly navigate from one segment to another. In HTML, the essential tags are the section headers, as marked up by \verb|<h1>...</h1>|,  \verb|<h2>...</h2>|, etc. GPT correctly interpreted
{\footnotesize
\begin{verbatim}
\begin{document}

\title{Testing \LaTeX{} Accessibility}
\author{Zaphod Beeblebrox}
 \affiliation{Sirius Cybernetics Corporation} 
\date{\today}

\begin{abstract}
This is a document to explore ...
\end{verbatim}
}

\noindent and inserted structures such as

{\footnotesize
\begin{verbatim}
<body>
    <header>
        <h1>Testing LaTeX Accessibility</h1>
        <p>Author: Zaphod Beeblebrox, Sirius ...</p>
        <p>Date: Today's Date</p>
    </header>
    <section>
        <h2>Abstract</h2>
        <p>This is a document to explore ... </p>
    </section>
\end{verbatim}
}

\noindent which not only use the correct section headers, but in one of the responses also the above shown additional semantic markup like the HTML5-tags \verb|<header>| and \verb|<section>|; those potentially helpful tags were missing in the second rendering. However, both versions were perfectly accessible to and navigable by screen-reader software.

\subsection{New Commands}
Several HTML-rendering tools fail to interpret \verb|\newcommand|. The test document included
{\footnotesize
\begin{verbatim}
\newcommand{\thesquareroot}{\sqrt{1-\frac{v^2}{c^2}}}
\end{verbatim}
}
\noindent which was used in nearly all subsequent formulas and interpreted correctly in every case. For example, \verb|$L'=L\thesquareroot$| was interpreted as ``L prime equals L times the square root of 1 minus v squared over c squared.''

\subsection{Equations}
GPT-4 overall did very well in narrating equations. As an example,
{\footnotesize
\begin{verbatim}
\begin{equation}
\oiint\vec{E}\cdot d\vec{S} =
   \frac{1}{\epsilon_0}\iiint\varrho dV
\end{equation}
\end{verbatim}
}
\noindent was translated into
\begin{quote}\textit{\noindent
The closed integral of E dot dS equals one over epsilon zero times the triple integral of rho dV.
}\end{quote}
\noindent in the first version (Fig.~\ref{fig:htmlversion1top}), and
\begin{quote}\textit{\noindent
The surface integral over the vector E dotted with the infinitesimal area vector dS equals 1 over the permittivity of free space times the volume integral over the volume charge density times the infinitesimal volume dV.
}\end{quote}
\indent in the second version (Fig.~\ref{fig:htmlversion2top}).

Both narratives are adequate and could be spoken by screen-reader software. The first rendering is rather literal, though, and fails to mention that the first integral is a surface integral, and that $\vec{E}$ and $d\vec{S}$ are vectors. A physicist will understand this concise narrative and probably be thankful that it is not too verbose, while a learner (actually, any learner, blind or seeing) would be thankful for the second, interpretive rendering. It is remarkable that GPT provided the physical meaning of the symbols alongside the narrative of their mathematical connections.

Another example is
{\footnotesize
\begin{verbatim}
\begin{equation}\label{eq:famous}
E=\frac{mc^2}{\thesquareroot}\stackrel{v=0}{=}mc^2
\end{equation}
\end{verbatim}
}
\noindent which is rendered
\begin{quote}\textit{\noindent
E equals mc squared divided by the square root of 1 minus v squared over c squared, which simplifies to mc squared when v equals 0.
}\end{quote}
\noindent in one version, and
\begin{quote}\textit{\noindent
E equals mc squared divided by the square root of (1 minus (v squared divided by c squared)) when v is not equal to 0. When v equals to 0, E equals mc squared.
}\end{quote}
\noindent in the other. Both are essentially equivalent, and both do a remarkable job of interpreting \verb|\stackrel| (even though, the full expressions is correct regardless of $v$ being zero or not).

A true challenge could have been posed by the rotation matrix:
{\footnotesize
\begin{verbatim}
\begin{eqnarray}
\vec{x}'&=&D\vec{x}\\
D&=&\left(\begin{array}{ccc}
\cos\theta&\sin\theta&0\\
-\sin\theta&\cos\theta&0\\
0&0&1
\end{array}\right)
\end{eqnarray}
\end{verbatim}
}
Intriguingly, this was incorporated into the running narrative in one version (Fig.~\ref{fig:htmlversion1bot}),
\begin{quote}\textit{\noindent
Consider a rotation matrix about the z-axis in three-dimensional space, x prime equals D x and D equals cos theta, sin theta, 0 in the first row, minus sin theta, cos theta, 0 in the second row, and 0, 0, 1 in the third row.
}\end{quote}
\noindent and spelled out rather literally in the other (Fig.~\ref{fig:htmlversion2bot}),

\begin{quote}\textit{\noindent
The vector x prime equals the matrix D times the vector x.\newline
The rotation matrix D is a 3x3 matrix with elements cos theta, sin theta, 0 on the first row, -sin theta, cos theta, 0 on the second row, and 0, 0, 1 on the third row.
}\end{quote}
\noindent Once again, both versions correctly describe the situation, and a blind person would be able to envision the matrix in his or her head based on the description.

On the other hand, the equation
{\footnotesize
\begin{verbatim}
\begin{equation}
x'^\nu=\sum_{\mu=0}^3\Lambda^\nu_{\ \mu}x^\mu
\end{equation}
\end{verbatim}
}
\noindent is the first example of a situation where one of the narratives falls short:
\begin{quote}\textit{\noindent
x prime nu equals the sum from mu equals 0 to 3 of Lambda nu mu x mu.
}\end{quote}
\noindent is less helpful than
\begin{quote}\textit{\noindent
x prime superscript nu equals the sum from mu equals 0 to 3 of lambda superscript nu subscript mu times x superscript mu.
}\end{quote}
\noindent since it fails to distinguish between covariant and contravariant indices. The nonchalant treatment of indices are also the reason for parenthetically introducing  ``including the indices'' into the prompt; without that phrase, the GPT-algorithm recognized the matrix multiplication for what it is and simply stated
\begin{quote}\textit{\noindent
The transformed x vector equals the sum from mu equals 0 to 3 of Lambda times the original x vector.
}\end{quote}
\noindent In fact, even before that, it already assumed the summation convention and rendered $x'^\nu=\Lambda^\nu_{\ \mu}x^\mu$ as
\begin{quote}\textit{\noindent
The transformed x vector equals Lambda times the original x vector.
}\end{quote}
\noindent This highly concise rendering would likely be appreciated by an expert, but might generally be too far from the original.

\subsection{Links and References}
Links and references were handled differently in both renderings. For example, toward the bottom of the LaTeX source, it says
{\footnotesize
\begin{verbatim}
One could say a lot more about Special Relativity,
but \textit{don't panic}~\cite{adams2010}!
At the very least, remember equation~\ref{eq:famous}!
\end{verbatim}
}
\noindent In the first version, the GPT algorithm had included
{\footnotesize
\begin{verbatim}
<p id="eq:famous">E equals mc squared divided ... </p>
\end{verbatim}
}
\noindent at the place where the formula appeared, as well as
{\footnotesize
\begin{verbatim}
<li id="adams2010">Adams, Douglas. The Ultimate ... </li>
\end{verbatim}
}
\noindent into the reference list, and it rendered the LaTeX as
{\footnotesize
\begin{verbatim}
... but <i>don't panic</i><a href="#adams2010">[5]</a>!
At the very least, remember equation
<a href="#eq:famous">2</a>!
\end{verbatim}
}
\noindent This in fact only works if following the hyperlinks, since neither the equation nor the reference have numbers appearing in the output (the reference list used the unordered list \verb|<ul>| instead of the ordered list \verb|<ol>|).  A blind user, however, could follow those hyperlinks to get to the correct equation and references.

The second version uses a different approach:
{\footnotesize
\begin{verbatim}
... but <em>don't panic</em>
<a href="#adams2010">Adams 2010</a>!
At the very least, remember equation
E equals mc squared divided by the square root
of (1 minus (v squared divided by c squared))!
\end{verbatim}
}
\noindent This approach works for both blind and seeing readers, but actually quoting the referenced equation could become awkward if the equation is a long one; in this case, GPT only quoted the first part, which may not have been what the author wanted to reference. This literature reference also works for seeing readers, since the author name is used. In this version, though, the table reference does not work for any readers, since at simply says
{\footnotesize
\begin{verbatim}
Table 1 lists some of the most commonly used ...
\end{verbatim}
}
\noindent but the table has no number.
\subsection{Tables}
The table in the source document was coded as
{\footnotesize
\begin{verbatim}
\begin{table}
\caption{\label{tab:formulas}Some relativistic formulas}
\begin{tabular}{p{1.8cm}p{4cm}l}
Common Name&Meaning&Formula\\\hline
Time\newline Dilation&Transformation ...
\end{tabular}
\end{table}
\end{verbatim}
}
\noindent which, as an example, in one the HTML versions, GPT interpreted as
{\footnotesize
\begin{verbatim}
<table id="tab:formulas">
        <caption>Some relativistic formulas</caption>
        <tr>
            <th>Common Name</th>
            <th>Meaning</th>
            <th>Formula</th>
        </tr>
        <tr>
            <td>Time Dilation</td>
            <td>Transformation ...
</table>
\end{verbatim}
} 

Both versions generated by GPT were WCAG compliant and can be navigated by screen readers. Particularly helpful is that the first row was interpreted as column headers (\verb|<th>| instead of \verb|<td>|), which is not even specified in the LaTeX markup.

\subsection{Figures}
While GPT-4 itself is multimodal, that is, it can support audio, image, and video input and output, the ChatGPT interface only allows for text-based conversations. However, the figure in the test document is generated using the popular Tikz package~\cite{walczak2008}:
{\footnotesize
\begin{verbatim}
\begin{figure}
\begin{tikzpicture}
\draw[thick,->] (2,0) -- (4,0)
           node[anchor=north west] {$x$};
\draw[thick,->] (2,0) -- (2,2)
           node[anchor=south east] {$y$};
\draw[dotted,thick,->] (2,0) -- (3.732,1)
           node[anchor=north west] {$x'$};
\draw[dotted,thick,->] (2,0) -- (1,1.732)
           node[anchor=south east] {$y'$};
\draw(3,0) node[anchor=south east] {$\theta$};
\end{tikzpicture}
\caption{Rotating a coordinate system.\label{fig:coord}}.
\end{figure}
\end{verbatim}
} 
GPT was prompted to describe this figure, but not draw it. The first version states
{\footnotesize
\begin{verbatim}
<figure>
    <figcaption>
         Figure 1: Rotating a coordinate system.
    </figcaption>
    <p>A coordinate system is represented
    with x and y axes.
    Two additional axes, x prime and y prime,
    are represented,
    rotated from the original x and y axes
    by an angle theta.</p>
</figure>
\end{verbatim}
}
\noindent while the second version has a simpler description that is woven into the narrative:
{\footnotesize
\begin{verbatim}
<p>This is illustrated in Figure 1 which shows a rotation
of a coordinate system with x, y axes and the rotated
x prime, y prime axes.</p>
\end{verbatim}
}
This second version correctly recognizes the contents of the figure, but fails to mention the angle $\theta$ that is also in the figure. Overall, both descriptions are useful.

\section{Discussion}
Ironically, if this paper gets published as a PDF, it will not be accessible to blind users. How can this be remedied, based on the findings of this exploratory study? In recent years, there has been a strong movement toward Open Access publication; based on this exploratory study, maybe publications should also become Open Source.

Over the coming years, it is to be expected that artificial-intelligence tools will increasingly become personalized agents; this is already suggested by the concepts that assistants like Siri and Alexa put forward, and by new product names like Copilot. In this study, the document rendered differently every time, and the prompt was fine-tuned in an attempt to avoid particular interpretations that experienced physicists might in fact prefer. One could argue that AI-assisted sense-making should be under the control of the user, and this means any user, having a disability or not. Thus, instead of mangling the document into a pure presentation format, the source code should be made available, so AI-tools can render it in ways preferred by the user.

This would mean free publication, not just in the sense of ``free beer,'' that is, free-of-charge, but in the sense of ``free speech,'' to reiterate an old free software adage~\cite{freesoftware}. Publishing the source code of manuscripts would likely have to go along with adequate licensing terms like Creative Commons~\cite{cc}. Publishers and preprint servers could make a corresponding link available to manuscripts published under those terms.

\section{Limitations}
The exact results of this study are not  reproducible, since the GPT-algorithm is not deterministic. Also, constructing a prompt for GPT is essentially trial-and-error, and better or more specific prompts than the one used for this study almost certainly exist. Finally, the study is purely exploratory: while the document was designed to be demanding in terms of LaTeX structures, there was no way to even remotely represent the creativity of LaTeX users. For large documents, GPT might encounter limitations due to its token limit.

\section{Conclusion}
Overall, GPT was found to be a useful tool to generate screen-reader compatible renderings of LaTeX documents, including document structure, equations, tables, references, and even figures that are based on textual source code. The output is compliant with accessibility standards.

As a probabilistic algorithm, the output is not reproducible, and it was found that each time a document is processed, some elements are translated better or worse than in other versions.
Notably, in situations where GPT is able to interpret equations based on its training data in physics, it is capable of generating narratives that are concise and desirable for experts, as well as more elaborative narratives that are suitable for learners.

\bibliography{LaTeXAccess}

\end{document}